OmniaScience

# Impact of family-friendly HRM policies in organizational performance

José María Biedma Ferrer, José Aurelio Medina Garrido

*Universidad de Cádiz (Spain)*

josemaria.biedma@uca.es, joseaurelio.medina@gm.uca.es

## Abstract

**Purpose:** The aim of this paper is to analyze the impact of family-friendly HRM policies in the perception of workers and their organizational performance.

**Design/methodology:** The authors have conducted a review of the major research related to work-family conflict and family-friendly HRM policies to overcome this conflict, to draw conclusions about its impact on worker performance.

**Findings:** This paper proposes an integrative model of the relationship between workfamily conflict and family-friendly HRM policies and perceptual variables on the one hand, and organizational behavior of the worker on the other hand. Perceptual variables analyzed are stress, job satisfaction and motivation. Regarding the variables related to organizational behavior worker absenteeism, rotations and performance were analyzed. The analysis results show evidence that the existence of work-family conflict influences negatively on the perceptual variables and organizational behavior of the worker, especially in performance. In turn, the existence of family-friendly HRM policies positively affects these variables.

**Originality/value:** This study integrates different perspectives related to work-family conflict and family-friendly HRM policies, from an eclectic vision. Thus, a more complete picture of the topic under investigation is provided. Furthermore, the proposed integrative model allows useful conclusions to management, both from the perspective of purely human resource management, and the improvement of organizational productivity.

## 1. Introduction

The incentive for many organizations to incorporate socially responsible measures could go beyond mere compliance with current regulations. In this sense, there are incentives to incorporate socially responsible measures that, at the same time, also mean an improvement in organizational performance. The implementation of work-life balance measures to reconcile work and family (hereinafter, work-life balance) is an intangible asset within corporate social responsibility, as it generates an improvement in the social image abroad. But, in addition, these conciliation measures could also imply an increase in the performance of workers. That is why the relationship between work-life balance and employee performance is a topic of great interest for organizations and, therefore, for academic research in the field of management.

Work-family conflict primarily occurs when work-related responsibilities interfere with family activities (Lin, 2013). Precisely, work-life balance measures present a wide variety of practices that help workers to balance the demands of their work and the obligations of their family life (Lobel & Kossek, 1996). It is clear that the difficulty of reconciling work with family responsibilities, as well as the increasing demand for results, have exerted strong pressures on employees (R. Burke, 2009; Rosso, 2008). In this sense, work-family conflict is a generator of stress that could have a negative impact on work performance and, consequently, on the organization's results (Beauregard & Henry, 2009; Yasbek , 2004).

It seems interesting that, from an organizational management perspective, this conflict is properly analyzed and managed. For this reason, some organizations have paid great attention to the work-family conflict and, in order to reduce it, have proposed measures to reconcile family life.

According to Beauregard and Henry (2009), there has been increasing pressure for the implementation of policies to support and improve the quality of life at work. In the same vein, it can be pointed out that research carried out on organisational practices that promote work-family balance shows, in general, a positive relationship between the reconciliation of family and work responsibilities and the improvement of employee performance in the companies that implement them (Clark, 2000; Hill, Hawkins, Ferris & Weitzman, 2001; Anderson, Coffey & Byerly, 2002).

The aim of this paper is to analyze the impact of implementing work-life balance measures on workers' perceptions and how this can influence their work behavior, especially their organizational performance.

The work is structured in six sections. After this first introductory section, the second section discusses the background and importance of the topic. The third section addresses the impact of conflict and work-family reconciliation on the worker and his or her behavior. In the fourth section, an integrative model is proposed that shows the relationships between conflict and work-family reconciliation and perceptual variables, on the one hand, and related to the worker's organizational behavior , on the other. The fifth section deals with discussion and recommendations for management. Finally, a series of conclusions are made from the work carried out.

## 2. Background and Importance of the Topic

From an academic point of view, there are several theories that deal with work-family conflict as a matter of research. From the point of view of business management, the theory of role, the theory of compensation and the theory of conflict are of interest.

Role Theory (Kahnm, Wolfe, Quin & Snoek, 1964) considers that the conflict between expectations associated with different roles has negative effects on personal well-being. This theory coincides with Karasek's (1979)  model of stress, where work-family conflict  acts as a stressor. There are numerous studies that examine Kahnm's theoretical model, linking antecedents, moderators, and consequences, in order to establish an integrative framework of the work-family relationship. Thismodel postulates social support, time commitment, and overload (both at work and in the family) as antecedents; as modulating variables, work-family conflict and family-work conflict; and as consequences, anxiety, dissatisfaction, and performance level.

Within this context, it is interesting to highlight the role of women. Traditionally, women have understood their private lives as a set of affective and material practices aimed at caring for and attending to others (Murillo, 1996). With modern society, there have been changes in the life trajectory of women that affect the sphere of work, family, etc. This process of change is irreversible, and a return to the traditional role of women is unthinkable (Beck-Gersheim, 2003).

On the other hand, the Compensation Theory considers that there is an inverse relationship between the work environment and the family, as well as between work activity and the fact of not working (Staines, 1980). This theory represents efforts to balance dissatisfaction in one domain by seeking satisfaction in the other (Edwards & Rothbard, 2000; Lambert, 1990). According to this theory, people will be involved differently in different environments, increasing their involvement in one at the expenseof the other, in such a way that an inverse relationship between work and family would be established. In this sense, Zedeck and Moiser (1990) point out that work and non-work experiences are often antithetical (antagonistic).

Conflict Theory (J.H. Greenhaus & Beutell, 1985) considers that job satisfaction requires sacrifice in family roles or vice versa. It is a type of inter-role conflict, whereby role pressures from either side are incompatible in some respect. Conflict is defined as an internal and close process of the subject.

Subsequently, these theories have been developed that, while acknowledging the existence of conflict, highlight the need to seek balance by facilitating the reconciliation between work and family (Grzywacz & Marks, 2000). In this way, the concept of Work-Family Balance has emerged. This concept has elements derived mainly from the notion of balance or non-balance between the work role and the family role (J. Greenhaus, Collins & Shaw 2003; J. Greenhaus & Powell, 2006; Kirchmeyer, 2000; Clark, 2000). On the other hand, several studies have been carried out to establish consequences for companies arising from the conflict vs. work-family reconciliation (Idrovo, 2006; Baxter & Chesters, 2011; Behan & Drobnic, 2010; Meil, García & Luque, 2008; Hammer, Bauer & Grandey, 2003; Wayne, Randel & Stevens, 2006; Carr, Boyar & Gregory, 2008; Wei-Chi, Chien-Cheng & Hui-Lu, 2007; Litchfield, Swanberg & Sigworth, 2004). In this sense, the effects that conflict and conciliation generate on the perception of workers and how they influence their behavior in the organization are analyzed below.

### 3. Organisational impact of conflict and work-life balance

In this section, the impact of work-family conflict, and the necessary work-life balance, on the perception and behaviour of the worker is studied. Thus, it will be analyzed, among other aspects, what is the relationship between conflict and work-family reconciliation and some factors that belong to the field of workers' perception, such as stress (Anderson et al., 2002), satisfaction (Bagger, Li & Utek, 2008) and motivation (López-Ibor, Ecot, Fernández & Palomo, 2010). Likewise, the impact on absenteeism (Anderson et al., 2002), abandonment of the organization (Drago, Costanza, Caplan, Brubaker, Cloud & Harris, 2001; Estes, Noonan & Maume, 2007) and worker performance (Swody & Powell, 2007).

### 3.1. Stress

Baxter and Chesters (2011) show how difficulty reconciling work and family life causes stress. Similarly, Zhang and Liu (2011) state that work-related stress is closely related to work-family conflict. Rodríguez and Nouvilas (2007) establish that the generic relationship between work and family becomes a conflict that, in turn, becomes a source of stress for individuals. Anderson et al. (2002) establish a direct and positive relationship betweenwork-family conflict and certain levels of work-related stress. Along these lines, Litchfield et al. (2004) pointed out that work-family balance policies have, among other benefits, the reduction of stress. There have also been some empirical studies that confirm that the lack of work-life balance measures generates stress (Idrovo, 2006).

### 3.2. Job satisfaction

Job satisfaction is defined as one's evaluation and feelings toward the job (Locke, 1976). Studies have found that people who enjoy a better work-family relationship are more likely to have greater job satisfaction (Aryee, Srinivas & Tan, 2005; Balmforth & Gardner, 2006).

Bagger et al. (2008), on the other hand, find a positive correlation between work-family conflict and job dissatisfaction, as well as a negative correlation between this type of conflict and job satisfaction. Similarly, Carr et al. (2008) analyze the negative impact of work-family conflict on job satisfaction. In the same vein, Anderson et al. (2002) establish an inverse relationship between work-family conflict and job satisfaction, which is exacerbated instressful situations. A high correlation between work-family balance and job satisfaction has been confirmed, establishing the importance of organizational and personal initiatives to increase satisfaction in the work-family relationship in employees (Behan & Drobnic, 2010). In the same vein, Kinnunen, Mauno, Geurts and Dikkers (2005) consider that those organisations that promote work-life balance have positive effects on specific aspects such asjob satisfaction.

Baltes, Briggs, Huff, Wright, and Neuman (1999) conducted a meta-analysis that showed evidence that both flexible work schedules and compressed workweeks had a positive effect on job satisfaction. T. Allen (2001) and Batt and Valcour (2003), on the other hand, also found that practices that involve work flexibility, Flexible working hours and compressed weeks are positively related to job satisfaction. In the same vein, Lingard and Francis (2006) state that flexible working hours are essential to alleviate work-family conflicts.

Employees' perception that the organization collaborates positively in the management of work-family conflict means that workers have a more favorable attitude and feelings about their work and the organization (Aryee et al., 2005; Wayne et al., 2006). This supportive attitude on the part of organizations also means that organizations that treat work-life balance favorably are more likely to be considered desirable by employees (Blair-Loy & Wharton, 2002; Kossek, Noe & Demarr, 1999; Leiter & Durup, 1996).

## 3.3. Statement of reasons

Research on work-family conflict does not pay enough attention to motivation (Zhang & Liu, 2011). However, it can be said that there is evidence about the relationship between this conflict and motivation. In this sense, López-Ibor et al. (2010) study how motivation increases with work-life balance measures. In the same vein, Meil et al. (2008) understand that motivation should be considered as one of the benefits derived from the application of conciliatory measures. Therefore, the two aspects have a clear relationship. On the other hand, Williams and Alliger (1999  ) considered that unpleasant moods can extend from work to family life and vice versa, leading to dysfunction of coping strategies often associated with *burn-out* (Leiter, 1991).

## 3.4. Absenteeism

It is considered common for absenteeism to be positively related to work-family conflict  (R.J. Burke & Greenglass, 1999). Anderson et al.'s (2002) model establishes a positive correlation between work-family conflict and family-work conflict. Considering that absenteeism from work maintains a positive relationship in the family-work conflict, absenteeism from work can be expected to increase when the work-family conflict increases. Studies conducted by Hammer et al. (2003) on work-family conflict show that absenteeism has significantnegative consequences for the firm. In this sense, some specific work-life balance measures are aimed at giving flexible working hours to workers so that they can attend to family or health issues that, in other circumstances, would generate absenteeism. Litchfield et al. (2004) have pointed out the decrease in absenteeism rates as a result of the implementation of work-life balance programmes.

## 3.5. Abandonment

Meyer and Allen (1997) define affective engagement as employees' emotional attachment and identification with and participation in the organization. Based on this concept, T.D. Allen, Herst, Bruck,  and Sutton (2000) demonstrate that work-family conflict is negatively associated with affective commitment. On the other hand, Kinnunen et al. (2005) point out that those organizations that favor work-life balance generate positive effects on organizational commitment. This is because employees' perception that the organization collaborates positively in the management of work-family conflict contributes to workers having a more favorable attitude and feelings about their work and the organization (Aryee et al., 2005; Wayne et al., 2006). In fact, there is empirical evidence that the lack of work-life balance measures leads to a lack of commitment and absenteeism (Idrovo, 2006).

Therefore, work-life conflict can have an impact on workers' desire to leave the organization (Boyar, Maertz, Alison, Pearson, & Keough, 2003; Batt & Valcour, 2003). In this sense, Anderson et al. (2002) consider that work-family conflict is positively related to the intention to leave the organization. That is why when corporations alleviate the tension between work and family, turnover is reduced  (Drago et al., 2001; Estes et al., 2007; Glass & Estes, 1997; Swody & Powell, 2007). Thus, it has been shown that company policies that favour the work-life balance of employees achieve greater worker retention (Foley, Linnehan, Greenhaus & Weer, 2006; Bocaz, 2003; Dex & Scheibl, 1999). Along these lines, Carr et al. (2008) analyze the negative impact of work-family conflict on organizational commitment and worker retention, among other issues. More specifically, Scandura and Lankau (1997) established that flexible working hours are associated with greater organizational commitment. In addition, T. Allen (2001) also found that practices that involve work flexibility are negatively related to turnover intentions. The variables they analyzed are availability of flexible schedules and a compressed workweek calendar.

## 3.6. Performance

The conflict between work and personal life results in lower organizational performance (Sánchez-Vidal, Cegarra-Leiva & Cegarra-Navarro, 2011). That is why when corporations relieve the tension between work and family, worker productivity increases (Drago et al, 2001; Estes et. al., 2007; Glass & Estes, 1997; Swody & Powell, 2007). Wei-Chi et al. (2007) empirically analyze how employees' positive mood predicts task performance indirectly through motivational (self-efficacy and task persistence) and interpersonal (helping and receiving help from peers) processes.

In this sense, the implementation of work-family policies is associated with high levels of organizational performance, increases in productivity, and an improvement in work motivation (Perry-Smith & Blum, 2000). Along these lines, research on organizational policies to promote work-family balance shows, in general, a positive relationship both in the reconciliation of work and family responsibilities, as well as better performance in the companies that operate them(Clark, 2000; Hill et al., 2001; Anderson et al., 2002). The academic literature shows empirical evidence of the relationship between work-life balance and increased labor productivity (Bocaz, 2003; Litchfield et al., 2004). Along these lines, Parasuraman  and Simmers (2001) empirically analyze the positive correlation between the stability of work shifts and productivity. On the other hand, there are empirical studies that analyse how the improvement in productivity allows the benefit obtained to be greater than the cost of implementing work-life balance measures (SERNAM, 2003).

### 4. Towards an integrative model

The work-family conflict, and the conciliation that can resolve this conflict, can be analysed as two extremes of the same dimension. In this sense, in order to propose an integrative model, both variables can be summarized in a single one, work-life balance  (see Figure 1). Where  the existence of work-life balance measures would mean a  positive value for this variable, and the existence of work-family conflict refers to the absence of work-life balance or a negative work-life balance.

The worker's perception of the support provided by the organization to reconcile work and family can have a positive impact on commitment to the organization, job satisfaction, employee retention, and performance (Dixon & Sagas, 2007), among other aspects related to the worker's perception and organizational behavior. As discussed in the previous section, several authors have considered work-family conflict to be astressful organizational variable that affects employee job satisfaction, absenteeism, and the intention to leave the organization (T.D. Allen et al., 2000). On the other hand, Anderson et al. (2002) propose a model that considers that work-family conflict is negatively related to job satisfaction and, on the contrary, is positively related to stress, the intention to leave the organization and, indirectly, to absenteeism from work.

Based on the analysis in the previous section, as well as on the contributions of the Anderson et al. (2002) model, Figure 1 describes a model in which the positive or negative relationships between work-life balance and, directly or indirectly, perceptual variables such as stress, satisfaction and motivation are reflected; and on variables related to worker behavior such as absenteeism, attrition and performance.

More specifically, Figure 1 shows how work-life balance is negatively related to stress, absenteeism, and job abandonment. On the other hand, work-life balance measures also have a positive influence on employee satisfaction, motivation and performance.

It should be noted that this theoretical model is a simplification of the relationships between the variables, in order to be able to carry out a subsequent empirical study of them, and to provide useful recommendations for human resources management. However, there are more complex relationships that have not been represented. For example, a negative relationship could be established between stress at work and satisfaction and motivation. At the same time, it could also be argued that lack of job satisfaction and motivationincreases the likelihood of absenteeism and job abandonment.

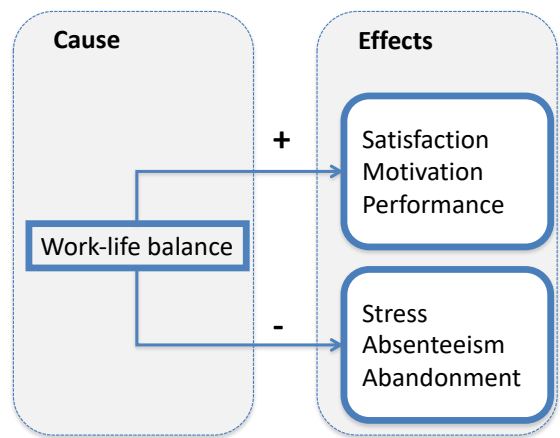

Figure 1. Influence of work-life balance on the perception and behaviour of the worker

In relation to the model in Figure 1, and in accordance with the literature reviewed in the previous section, the following working proposals can be proposed:

- Proposition 1: There is a positive and meaningful relationship between work-life balance and job satisfaction.

- Proposition 2: There is a positive and significant relationship between work-life balance and motivation at work.

- Proposition 3: There is a positive and meaningful relationship between work-life balance and job performance.

- Proposition 4: There is a negative and significant relationship between work-life balance and stress at work.

- Proposition 5: There is a significant negative relationship between work-life balance and absenteeism.

- Proposition 6: There is a negative and significant relationship between work-life balance and leaving the organization.

Therefore, work-family reconciliation measures are an element of conflict reduction. While conflict could generate negative perceptions and behaviors in the organization, work-life balance is positively correlated with the aforementioned perceptual and behavioral variables. In addition, this correlation positively influences, in turn, the performance of workers.

## 5. Discussion and future lines of research

There is evidence that companies have much to gain when they help workers in the pursuit of work-life balance (Lingard & Francis, 2006).

Work-life balance measures can be an important tool in human resource management both to reduce negative perceptual factors and to promote desirable behaviors. Among the perceptual factors to be reduced are stress, lack of satisfaction and motivation. As for the behaviors that can be positively influenced by implementing work-life balance measures, there would be a reduction in absenteeism, abandonment of the organization and improvements in performance.

In this context, it is worth analysing the work-life balance measures that will improve the impacts mentioned above. In this sense, it is appropriate to carry out an empirical analysis that detects the greatest statistical correlations between specific measures of work-life balance and the previous perceptual and behavioural variables. In addition, it is interesting to analyse significant differences in the above correlations according to the sector in which one works, sex, age and other socio-demographic variables. This analysis will enable management to make better decisions from allocating limited resources to the most efficient conciliatory measures. In addition, it will make it possible to improve the corporate image transmitted to society by its appropriate socially responsible human resources practices.

## 6. Conclusions

The interest of many organizations in incorporating socially responsible measures does not have to be incompatible with seeking an improvement in organizational performance. In this sense, according to the review of the literature carried out in this work, it is observed that the implementation of work-life balance measures is an intangible capital within corporate social responsibility that, in addition, could imply an increase in the performance of workers.

This paper proposes an integrative model that shows the relationships between work-family reconciliation (or, in its absence, work-family conflict) and some perceptual variables of the individual, on the one hand, and variables related to the organizational behavior of the worker, on the other. Among the perceptual variables, stress, job satisfaction and motivation were analyzed. Regarding the variables related to the organizational behavior of the worker,absenteeism, abandonment and performance were analyzed. The results of the analysis show evidence that the existence of work-family balance is perceived favorably by workers and improves their organizational behavior, especially their performance.

This study provides the existing literature with an integrative view of the topic analyzed, which has traditionally been studied in a fragmented way. In addition, the integrative model that is proposed makes it possible to extract useful recommendations for management, both from a purely human resources management perspective and from the improvement of organizational productivity. Work-family reconciliation measures are an element of conflict reduction.
These measures could generate favorable perceptions and behaviors in the organization.

Finally, in the discussion section, future lines of scientific research have been proposed that can be developed from the variables raised in this work. In this sense, it is appropriate to carry out a subsequent empirical study that analyzes the statistical correlations between specific measures of reconciliation and the previous perceptual and behavioral variables. In addition, it is interesting to analyse significant differences in the above correlations according to the sector in whichone works, sex, age and other socio-demographic variables.